# Investigation of the characteristics of the electromagnetic induction transparent-like spectrum with counter-propagating waves coupling mechanism for waveguide and micro-ring coupled system

Chaoying Zhao[a,b]*

[a]Hangzhou Dianzi University, College of Science, Zhejiang, China [b]Shanxi University, State Key Laboratory of Quantum Optics and Quantum Optics Devices, Taiyuan, China

E-mail: zchy49@163.com

**Abstract**

  In this paper, a new counter-propagation waves coupling mechanism is proposed which is expected to realize an electromagnetically induced transparency (EIT)-like effect. Comparing the traveling waves coupling mechanism (see *J. Mod. Opt.* 2015,62:313-320 [9]) with the counter-propagating waves coupling mechanism, we find out that the transparency window breadths of transmission spectra are greatly enhanced and the corresponding phase shift spectra possess a flat profile or a square profile. Our numerical simulated results are in good agreement with the theoretical analysis. The EIT-like effect can significantly reduce the group velocity near the edge of the square profile transparent window. We believe that the counter-propagating waves coupling mechanism is particularly beneficial for the realization of active manipulation of slow light devices (such as delay lines) required in the conventional EIT scheme. In the vicinity of the transparency peak, we can obtain a large group delay, may gain more significant potential applications in slow-light transmission and optical storage.

## 1. Introduction

The Electromagnetic induced transparency (EIT) effect occurs when the atomic medium accompany a strong dispersion, which can significantly reduce the group velocity of light.

---

* Author to whom any correspondence should be addressed.

EIT is widely used in slow light[1,2], optical delay lines[3], optical storage[4] and low-loss optical devices[5]so on. However, the experimental realization of EIT effect in atomic system usually requires some complicated conditions, such as temperature, high intensity laser, etc. Therefore, the application of EIT effect in practice has been greatly limited. The all-optical analog to EIT based on wave-guide and micro-ring coupled system has attracted much more attention in recent years. The EIT-like phenomena have been theoretically analyzed in wave-guide coupled micro-ring[6], and wave-guide coupled two micro-ring[7]. Meng *et.al* have studied the transmission spectra of $N \times N$ weak linearly array[8]. We have investigated the wave-guide coupled double-micro-ring systems by using the traveling wave coupling mechanism[9]. In 2008, Zhang *et. al.* first proposed the concept of PIT (plasmon-induced transparency) in metamaterials[10], Metamaterials are a class of artificial electromagnetic media, aiming to provide some controllable electromagnetic properties, such as room temperature conditions and large operating bandwidth. They pointed out that the EIT-like effect can be simulated by adding a special resonance structure (dark state resonance unit). EIT-like effect can be explained by the near-field coupling principle between bright and dark resonant units. In 2021, Zhao et.al have studied dual-band electromagnetically induced transparency (EIT)-like effect in terahertz spectrum. The EIT-like effect can significantly reduce the group speed near the transparent window. The EIT-like effect can significantly reduce the group speed near the transparent window, may gain more sigificant potential applications in slow-light transmission and optical storage [11]. In this paper, we want to investigate the transmission spectra and phase shift by using the counter-propagating waves coupling mechanism. We find out that the transparency window breadths of transmission spectra are greatly enhanced when the traveling wave coupling is substituted by counter-propagating waves coupling, and the phase shift spectra possess a flat or square profile, which result in group velocity slowdown very large at the edge of the square profile. This paper is organized as follows. In Section 2, the transmission of $2 \times 2$ coupled system for the traveling wave coupling and the counter-propagating waves coupling is studied in Section 3. The transmission and phase characteristic of $2 \times 2$ and $3 \times 3$ coupled systems containing single micro-ring, doule micro-ring (which is denoted by 'Sm' in the following) are calculated and analyzed in Section 4, and in Section 5, the conclusions are given.

**2.The transmission of traveling wave coupling and counter-propagating waves**

## coupling for $2 \times 2$ system

For $2 \times 2$ system, Fig.1(a) depicts the schematic structure of the micro-ring resonator, which consists of straight wave-guide $a_2b_2$, ring $a_1b_1$. The optical field from a excitation wave-guide passing through the evanescent tail couples into ring. After propagating a round trip, the wave couples back to the excitation wave-guide and interferes with transmitted light. At resonance, the wave appears destructive interference. The propagation path of light presents 'clockwise direction' in ring (see Fig.1(a)). As shown in Figure 1(a), the counter-clockwise mode $a_1 = \alpha e^{j\theta} b_1$ in the micro-ring $a_1 b_1$, is induced by the traveling wave $\varepsilon = \varepsilon_0 e^{-j(\omega t - kz)}$ in the waveguide $a_2 b_2$ from the left to the right, where $\alpha$ is the absorption coefficient, $\theta = \omega L/c$ is the round trip phase shift, $L$ represents the circumference of the micro-ring $a_1 b_1$, $c$ is the phase velocity of micro-ring mode, $\omega$ is the optical wave oscillation frequency.

While the light circuits clockwise and counterclockwise direction in ring simultaneously (see Fig.1(b)). We can analyze the gap parameter makes an influence on the EIT-like spectrum. For Figure 1.(b), the clockwise mode $a_1 = \alpha e^{-j\theta} b_1 /2$ in the micro-ring $a_1 b_1$, which is induced by the traveling wave $\varepsilon = \varepsilon_0 e^{-j(\omega t + kz)}$ in the waveguide $a_2 b_2$ from the left to the right. The counter-clockwise mode $a_1 = \alpha e^{j\theta} b_1 /2$ in the micro-ring $a_1 b_1$, which is induced by the traveling wave $\varepsilon = \varepsilon_0 e^{-j(\omega t - kz)}$ in the waveguide $a_2 b_2$ from the right to the left.

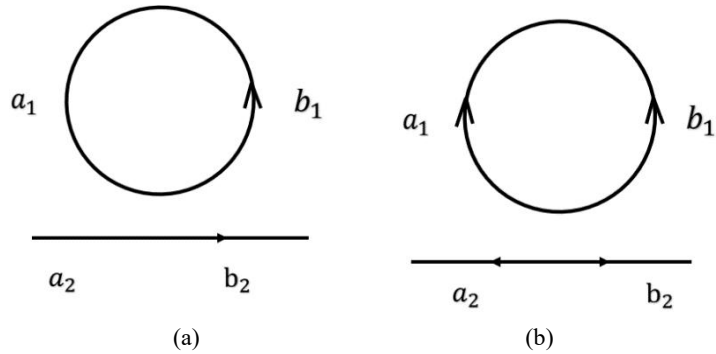

(a)      (b)

**Figure 1** Schematic diagram of $2 \times 2$ system, (a) traveling wave coupling, (b) counter-propagating wave coupling.

Assuming that the light wave is input from the straight waveguide input, when the resonance condition is satisfied, the phase before the light is coupled from the straight waveguide to the microring is $\Phi_1 = \varphi$. According to the optical waveguide coupling theory, the changes in the phase before and after a light wave is coupled from one waveguide to another. Then, when the light wave is coupled from the straight waveguide to the micro-ring waveguide, its phase becomes $\Phi_2 = \varphi + \pi/2$. According to the resonance relationship, when the light wave meets the resonance condition, the light circles in the microring and interferes with each other to generate resonance enhancement. The change $2m\pi$ of the phase before and after the light wave circulates in the microring, where the phase is $\Phi_3 = \varphi + \pi/2 + 2m\pi$. Similarly, the light wave is again coupled from the microring to the straight waveguide, and its phase changes again $\pi/2$, becoming $\Phi_4 = \varphi + \pi + 2m\pi$. Comparing the phase changes at the input and output, we find an $\pi$ odd multiple of the difference between $\Phi_1$ and $\Phi_4$, so that the two beams interfere and cancel. When the intensity of the light field is the same, the intensity of the light field after interference elimination is zero, so the output end of the straight waveguide is zero.

## 3. The influence of counter-propagation optical waves with phase difference

Figure 2 show the Schematic diagram to produce the counter-propagating waves.

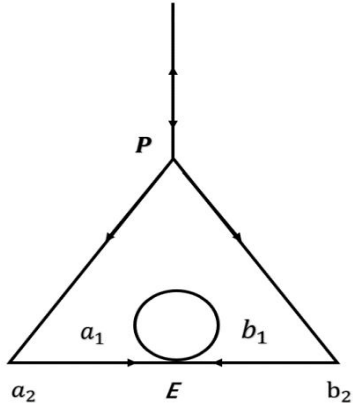

**Figure 2** Schematic diagram producing the counter-propagating waves.

As shown in Figure 2, one traveling wave is split into two equal portions of counter-clockwise wave and clockwise wave at the point *P*. Therefore, the superposed induced counter-clockwise wave and clockwise wave of the micro-ring assuming $(e^{j\theta} + e^{-j\theta})/2 = \cos[\theta]$. Taking account the absorption coefficient $\alpha$ around the micro-ring, we have the relation $a_1 = \alpha \cos[\theta] b_1$ for counter-propagating waves, the mode $a_1 = \alpha (e^{j\theta} + e^{-j\theta}) b_1 /2 = \alpha \cos[\theta] b_1$ in the micro-ring is induced by the counter-propagating waves $\varepsilon = \varepsilon_0 (e^{-j(\omega t - kz)} + e^{-j(\omega t + kz)})/2$ in the waveguide $a_2 b_2$.

For the traveling wave propagating in waveguide[4], the relation $a_1 = \alpha e^{j\theta} b_1$ is also valid. In general, the coupling system contains one waveguide and multi-micro-rings [7,9].

In the derivation of traveling waves coupling and counter-propagation waves coupling formula $e^{-j\theta} \to \cos[\theta]$, we have assumed that the path from $p$ to $E$ (see Figure 2) along the clockwise and the counter-clockwise waves have the same length. In general, this lengths may be different, the initial phases at point $E$ being $\theta_1$, $\theta_2$ for the clockwise and the counter-clockwise waves, respectively. The superposed induced waves being

$$\frac{1}{2}(e^{-j(\theta-\theta_1)} + e^{j(\theta+\theta_2)}) = \frac{1}{2}(e^{-j(\theta-\frac{\theta_1-\theta_2}{2})} + e^{j(\theta-\frac{\theta_1-\theta_2}{2})})e^{j\frac{\theta_1+\theta_2}{2}} = \cos[\theta - \frac{\theta_1-\theta_2}{2}]e^{j\frac{\theta_1+\theta_2}{2}}, (1)$$

Therefore, the traveling waves coupling and counter-propagating waves coupling may be represented by $\cos[\theta - \delta_1]e^{-j\delta_2}$, $\delta_1 = (\theta_1 - \theta_2)/2$, $\delta_2 = (\theta_1 + \theta_2)/2$, respectively. In the case of $\delta_1 = \theta$, $\cos[\theta - \delta_1]e^{-j\delta_2} = e^{-j\delta_2}$ possesses the form similar to that of the traveling waves coupling.

## 4. The numerical calculations and analysis of the transmission and phase shift spectra for $2 \times 2$ system and $3 \times 3$ system

In this section, we calculate the transmission and phase shift spectra reduction near resonance concern with the traveling waves coupling and the counter-propagating waves coupling.

In the following, we begin with the traveling wave coupling of $2 \times 2$ system. The transmission and the phase shift spectra [5]

$$T^{trl}_{2\times 2} = \left|\frac{t - \alpha e^{i\theta}}{1 - t\alpha e^{i\theta}}\right|^2, \quad \Phi^{trl}_{2\times 2} = Arg[\frac{t - \alpha e^{i\theta}}{1 - t\alpha e^{i\theta}}], \quad (2)$$

Figure 3 shows the transmission and the phase shift spectra for $T^{trl}_{2\times 2}$, $\Phi^{trl}_{2\times 2}$, respectively.

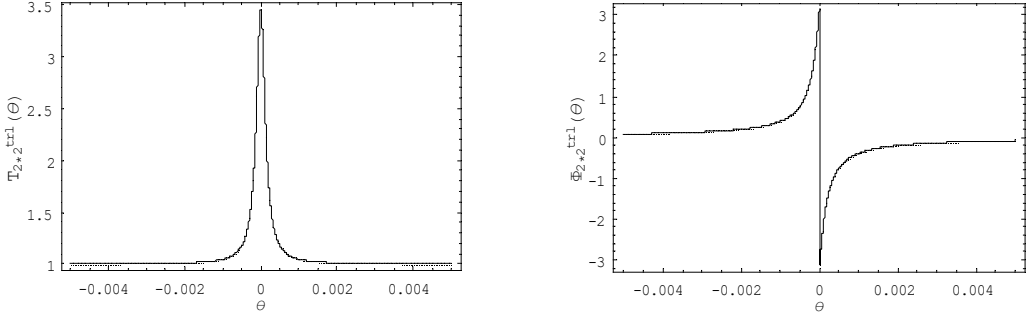

**Figure 3** The transmission and phase shift spectra with the parameters $t = 0.9998$, $\alpha = 1.00006$ versus $\theta$ in unit of radian.

And the corresponding $T_{2\times2}^{cou}$, $\Phi_{2\times2}^{cou}$ for counter-propagating wave coupling being

$$T_{2\times2}^{cou} = \left|\frac{t - \alpha \cos[\theta]}{1 - t\alpha \cos[\theta]}\right|^2, \qquad \Phi_{2\times2}^{cou} = Arg[\frac{t - \alpha \cos[\theta]}{1 - t\alpha \cos[\theta]}], \qquad (3)$$

Figure 4 shows the transmission and the phase shift spectra for $T_{2\times2}^{cou}$, $\Phi_{2\times2}^{cou}$, respectively.

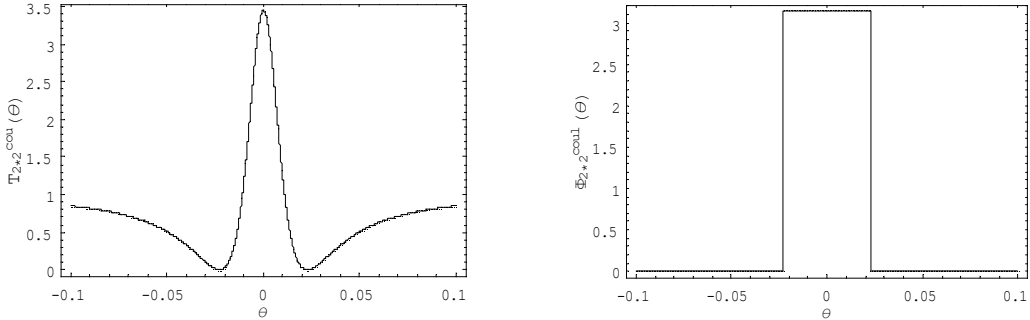

**Figure 4** The transmission and phase shift spectra with the parameters $t = 0.9998$, $\alpha = 1.00006$ versus $\theta$ in unit of radian.

Similarly, the transmission and phase shift spectra for the waveguide and two micro-rings coupled system [7]

$$\tau_2(\phi_1, \phi_2) = \frac{r_2 - a_2\, \tau_1(\phi_1)e^{i\phi_2}}{1 - r_2 a_2\, \tau_1(\phi_1)e^{i\phi_2}}, \qquad \tau_1(\phi_1) = \frac{r_1 - a_1 e^{i\phi_1}}{1 - r_1 a_1 e^{i\phi_1}}, \qquad (4)$$

Setting $\phi_1 = \phi_2 = \phi$, we have

$$T_{sm}^{trl} = |\tau_2(\phi, \phi)|^2, \qquad \Phi_{sm}^{trl} = Arg[\tau_2(\phi, \phi)], \qquad (5)$$

Figure 5 gives out the transmission and the phase shift spectra for $T_{2\times2}^{trl}$, $\Phi_{2\times2}^{trl}$, respectively.

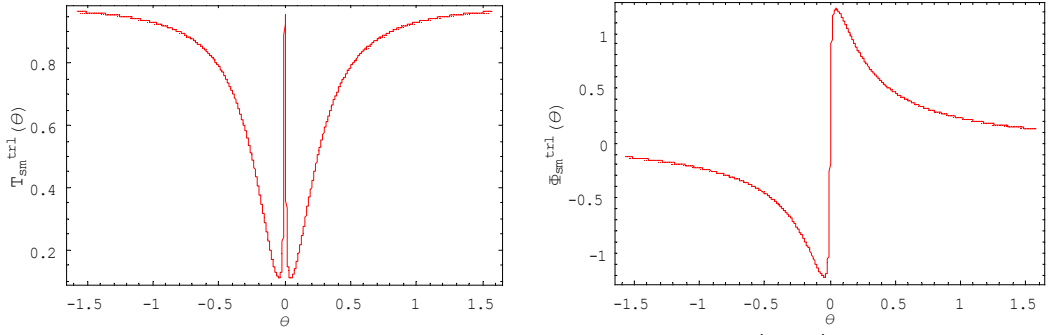

**Figure 5** The transmission and the phase shift spectra for $T_{sm}^{trl}$, $\Phi_{sm}^{trl}$ with parameters $r_1 = 0.999$, $r_2 = a_1 \times a_2$, $a_1 = 0.9999$, $a_2 = 0.88$ versus $\theta$ in unit of radian.

The corresponding formula for counter-propagating waves are

$$\tau_2^{cou}(\phi_1,\phi_2) = \frac{r_2 - a_2 \tau_1^{cou}(\phi_1)\cos[\phi_2]}{1 - r_2 a_2 \tau_1^{cou}(\phi_1)\cos[\phi_2]}, \quad \tau_1^{cou}(\phi_1) = \frac{r_1 - a_1 \cos[\phi_1]}{1 - r_1 a_1 \cos[\phi_1]}, \qquad (6)$$

Setting $\phi_1 = \phi_2 = \phi$, we have

$$T_{sm}^{cou} = \left|\tau_2^{cou}(\phi,\phi)\right|^2, \quad \Phi_{sm}^{cou} = Arg[\tau_2^{cou}(\phi,\phi)], \qquad (7)$$

Figure 6 gives out the transmission spectra and the phase shift spectra for $T_{2\times 2}^{cou}$, $\Phi_{2\times 2}^{cou}$, respectively.

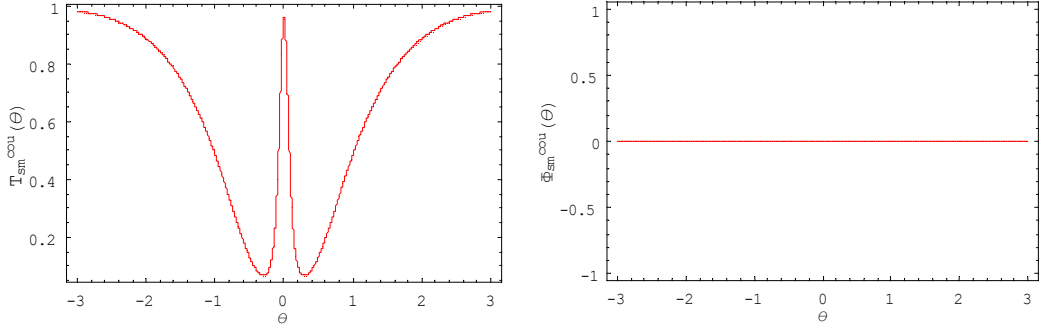

**Figure 6** The transmission and phase shift spectra for $T_{sm}^{cou}$, $\Phi_{sm}^{cou}$ with the parameters $r_1 = 0.999$, $r_2 = a_1 \times a_2$, $a_1 = 0.9999$, $a_2 = 0.88$ versus $\theta$ in unit of radian.

Figure 7 shows the schematic diagram of $3 \times 3$ system, (a) the traveling wave coupling, (b) the counter-propagation waves coupling.

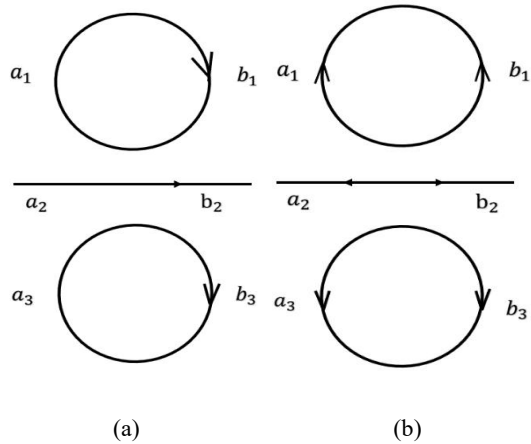

(a)　　　　　　　　(b)

**Figure 7.** Schematic diagram of $3\times 3$ system, (a)the traveling wave coupling, (b)the counter-propagation waves coupling.

The transmission and the phase shift spectra for $3\times 3$ system[12].

$$T_{3\times 3}^{trl} = \left|\frac{t - \mu B_1 - \delta B_3 + B_1 B_3}{1 - \delta B_1 - \mu B_3 + t B_1 B_3}\right|^2, \quad B_1 = \alpha e^{i\theta}, \quad B_3 = \alpha e^{-i\theta},$$

$$\Phi_{3\times 3}^{trl} = Arg[\frac{t - \mu B_1 - \delta B_3 + B_1 B_3}{1 - \delta B_1 - \mu B_3 + t B_1 B_3}], \tag{8}$$

Figure 8 gives out the transmission and the phase shift spectra for $T_{3\times 3}^{trl}, \Phi_{3\times 3}^{trl}$, respectively.

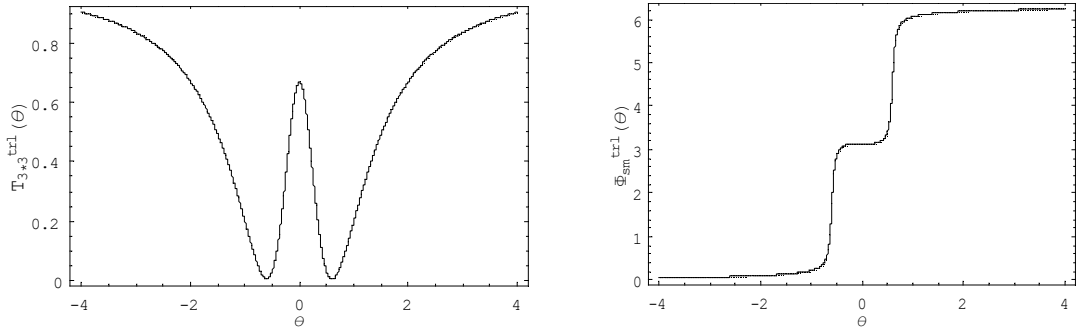

**Figure 8** The transmission and phase shift spectra for $T_{3\times 3}^{trl}, \Phi_{3\times 3}^{trl}$ with parameters $t = 0.9998, \alpha = 0.99998, \mu = 0, \delta = 0$ versus $\theta$ in unit of $10^{-4}$ radian

The transmission and the phase shift spectra for $3\times 3$ system[9].

$$T_{3\times 3}^{cou} = \left|\frac{t - \mu B_1 - \delta B_3 + B_1 B_3}{1 - \delta B_1 - \mu B_3 + t B_1 B_3}\right|^2, \quad B_1 = \alpha \cos[\theta], \quad B_3 = \alpha \cos[\theta],$$

$$\Phi_{3\times 3}^{cou} = Arg[\frac{t - \mu B_1 - \delta B_3 + B_1 B_3}{1 - \delta B_1 - \mu B_3 + t B_1 B_3}], \tag{9}$$

Figure 9 gives out the transmission and the phase shift spectra for $T_{3\times 3}^{cou}, \Phi_{3\times 3}^{cou}$, respectively.

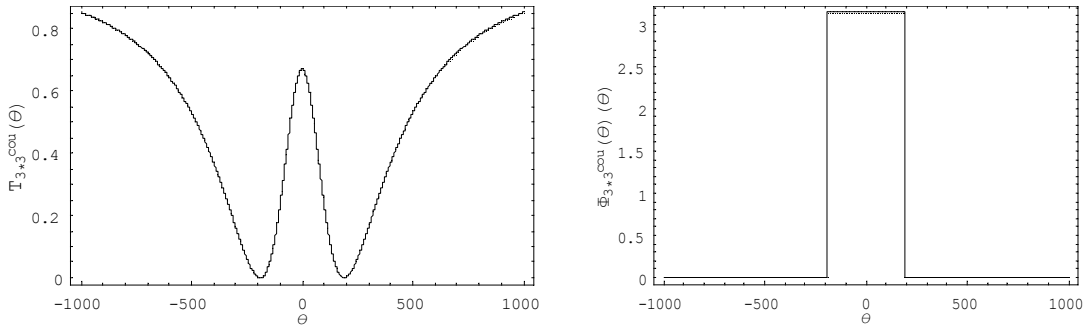

**Figure 9** The transmission and phase shift spectra for $T_{3\times 3}^{cou}, \Phi_{3\times 3}^{cou}$ with parameters $t = 0.9998, \alpha = 0.99998, \mu = 0, \delta = 0$ versus $\theta$ in unit of $10^{-4}$ radian.

The numerical calculations shown in Figures.3-6, 8-9, refer to that of the traveling wave coupling, the breadths of the transmission spectra for counter-propagating waves coupling are enhanced by a factor of $0.02/0.001 = 20$, $0.3/0.05 = 6$, $200/0.8 = 250$ for the $2\times 2$, $Sm$, $3\times 3$ system, respectively.

## 5. The analysis of the phase velocity spectra

We have obtained the phase spectra $\Phi = \Phi(\omega)$, in the case of counter-propagating waves coupling, the phase shift spectra possess a flat ($\Phi(\omega) = \Phi_{sm}^{cou}$) or a square profile ($\Phi(\omega) = \Phi_{2\times 2}^{cou}, \Phi_{3\times 3}^{cou}$). The inverse of the phase velocity is given by $c/v_g = n + d\Phi_{3\times 3}^{cou}/d\theta = n + c\,d\Phi_{3\times 3}^{cou}/L d\omega = n + c\,\tau_g/L$ [12], where $\theta = \omega L/c$, the group delay $\tau_g = d\Phi_{3\times 3}^{cou}/d\omega$ represents the time delay of narrow-band optical pulses in optical devices. The strong phase dispersion around the transparent window can cause a large group delay.

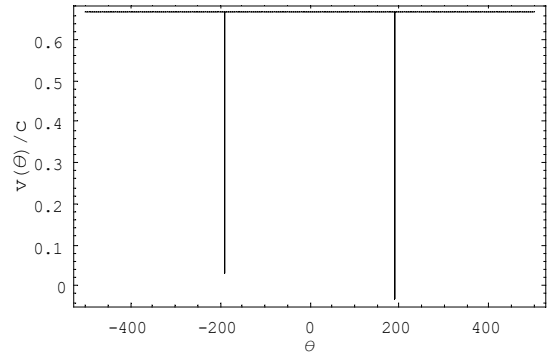

**Figure 10**. The phase velocity spectra for $3\times 3$ system versus $\theta$ in unit of $10^{-4}$ radian.

As shown in Figure 10, $v_g$ approaches to zero (namely a large group delays $\tau_g$ are obtained) at the edge of the square profile (namely in the vicinity of the transparency peak, see Figure 9).

The performance characteristics of micro-ring resonator are compared in Table 1.

Table 1 Comparison of EIT-like characteristic parameters based on different resonator

| Micro-ring resonator | microcavity structure | Number of EIT | coupling mechanism | material | group delays |
|---|---|---|---|---|---|
| [9] | two micro-ring | single | traveling waves | silica | --- |
| [11] | multiple U-shaped resonators | multiple | dark-light /light-light mode | metamaterials | high |
| [13] | chaotic single micro-ring | single | chaos-assisted dynamical tunneling | silica | --- |
| [14] | two microtoroid | single | traveling waves | silica | --- |
| Our work | single/two | single | counter-propagating waves | silica | high |

# 6. Loss analysis

According to the coupling mode theory, optical loss can directly affect the line type and characteristics of EIT spectrum of the system. For the dual-coupling microring resonator proposed in this paper, the transmission loss is mainly composed of scattering loss $1/Q_{scat}$ caused by the roughness of the waveguide side wall, bending loss $1/Q_{bend}$ of the microring waveguide and leakage loss $1/Q_{leak}$ caused by the light field entering the substrate. $1/Q_{mat}$ is the absorption loss of microring resonator. As the light waves travel in the waveguide, part of the light field is absorbed by the silicon material. In addition, the top layer of the optical waveguide device also has an oxide coating, which can absorb the light field in different environments. Finite element analysis (FEM) was used to conduct mode analysis on the waveguide cross section, and negative imaginary part ($k$) was added to the material refraction to obtain the absorption loss of the whole system [1]. $1/Q_{leak}$ is leakage loss, which is an important loss mechanism in SOI structure, and is generated by the middle spectral field of the waveguide into the silicon dioxide layer and silicon substrate. The leakage loss decreases with the increase of silica layer thickness. $1/Q_{scat}$ is the scattering loss, which occurs when the side wall of the waveguide is rough. Increasing the ratio of the width to the height of the waveguide, reducing the mode overlap at the side of the waveguide, and improving the fabrication technology of the waveguide can reduce the scattering loss of the system. In practical optical waveguide devices, the roughness of the side wall is larger than that of the surface. $1/Q_{bend}$ is the bending loss, is the radiation loss caused by the bending of the optical waveguide, and decreases with the increase of the radius of the microring. When the radius of the microring is about $30\mu m$ [15], the bending loss is about $0.015 dB/180$ [16]. $1/Q_{back}$ represents backscattering loss, which is one of the loss sources in waveguides based on SOI structure. Both the roughness of the waveguide side wall and the directional coupling can cause backscattering losses, which can be suppressed by various ways, such as improving the etching process to reduce the roughness of the waveguide surface, and lower $Q$ values [17]. In this paper, except for the feedback arm region, the coupling distance of the microring resonator is $0.15\mu m$. At this time, the resonance splitting has little influence on the transmission spectrum, so the backscattering can be ignored.

Assuming that the transmission loss of microring resonator mainly includes absorption loss ($1/Q_{mat}$), leakage loss ($1/Q_{leak}$) and bending loss ($1/Q_{bend}$). Therefore, the relationship between the value of $Q_{tot}$ the system and the loss coefficient $a$ can be expressed as [18]:

$$Q = \frac{2\pi n_g}{\lambda a} \quad (10)$$

$$a = \exp(-\frac{\alpha L_{ring}}{2}) \quad (11)$$

Where, $n_g$ is the group refractive index, which can be obtained by the finite-difference time-domain method of full-vector calculation by MODE Solutions software. $a$ is the round-trip loss coefficient of light wave in microring waveguide, $\alpha$ represents the loss value per unit length of optical waveguide, and $Q_{tot}$ is inversely proportional to $\alpha$. Taking all loss mechanisms into consideration, the transmission loss of the optical waveguide in this paper is about $9dB/cm$ [19].

## 7.Conclusions

In conclusion, we have demonstrated the transmission spectra and the phase shift spectra of two micro-rings coupled system through numerical simulations and theoretical calculations and analyzed the causes of resonance peaks and transparent windows leads to the EIT-like effects. As we can see in this paper, we can generate a wide transparent windows, a large flat phase shift and a large group delay by increasing the number of micro-ring resonators. Therefore, it can be potential applied in multi-band filters and multi-band slow light devices (such as delay lines) in optical communication field.

## Acknowledgments

This work was supported by the State Key Laboratory of Quantum Optics and Quantum Optics Devices, Shanxi University, Shanxi, China (KF202004，KF202205).

ORCID iD

Chaoying Zhao    https://orcid.org/0000-0003-1116-0790

## References

[1] Longdell J J, Fraval E, Sellars M J and Manson N B 2005 Stopped light with storage times greater than one second using electromagnetically induced transparency in a solid *Phys Rev Lett* **95** 063601

[2] Jahromi M A F and Bananej A 2016 Tunable slow light in 1-D photonic crystal *Optik* **127** 3889-91

[3] Safavi-Naeini A H, Mayer Alegre T P, Chan J, Eichenfield M, Winger M, Lin Q, Hill J T, Chang D E and Painter O 2011 Electromagnetically induced transparency and slow light with optomechanics *Nature* **472** 69-73

[4] Liu C, Dutton Z, Behroozi C H and Hau L V 2001 Observation of coherent optical information storage in an atomic medium using halted light pulses *Nature* **409** 490-3

[5] Zhu L, Dong L, Guo J, Meng F-Y, He X J, Zhao C H and Wu Q 2017 A low-loss electromagnetically induced transparency (EIT) metamaterial based on coupling between electric and toroidal dipoles *RSC Advances* **7** 55897-904

[6] Yariv A 2000 Universal relations for coupling of optical power between microresonators and dielectric waveguides. *Electron. Lett.* **36** 321-322

[7] Smith D D, Chang H 2004 Coherence phenomena in coupled optical resonators *J. Mod. Opt.* **51** 2503-2513

[8] Meng Y C, Guo Q Z, Tan W H, Huang Z M 2004 Analytical solutions of coupled-mode equations for multiwaveguide systems, obtained by use of Chebyshev and generalized Chebyshev polynomials *J. Opt. Soc. Am. A* **21** 1518-1528

[9] Zhao C Y, Tan W H 2015 Transmission of asymmetric coupling double-ring resonator *J. Mod. Opt.* **62** 313-320

[10] Zhang S, Genov D A, Wang Y, Liu M and Zhang X 2008 Plasmon-induced transparency in metamaterials *Phys Rev Lett* **101** 047401

[11] Zhao C Y, Hu J H 2021 Investigation of the characteristics of the dual-band electromagnetic induction transparent-like terahertz spectrum in a grating-like structure *J. Opt.* **23** 115103

[12] Heebner J E, Boyd R W, Park Q H 2002 Slow light, induced dispersion, enhanced nonlinearity, and optical solitons in a resonator-array waveguide *Phys. Rev. E* **65** 036619

[13] Xiao Y F, Jiang X F, Yang Q F, Wang L, Shi K B, Li Y, Gong Q H 2013 Tunneling-induced transparency in a chaotic microcavity *Laser Photon. Rev.* **7** L51-L54

[14] Zheng C, Jiang X S, Hua S Y, Chang L, Li G Y, Fan H B, Xiao M 2012 Controllable optical analog to electromagnetically induced transparency in coupled high-Q microtoroid cavities *Opt. Exp.* **20** 18319-18325

[15] Gondarenko A, Levy J S, Lipson M 2009 High confinement micron-scale silicon nitride high Q ring resonator[J]. *Opt. Exp.* **17** 11366-11370

[16] Chen J, Xie J, Wu K, et al. 2017 Continuously tunable ultra-thin silicon waveguide optical delay line *Optica* **4** 507-515.

[17] Li A, Van Vaerenbergh T, De Heyn P, et al. 2016 Backscattering in silicon microring resonators: a quantitative analysis *Laser & Photon. Rev.* **10** 420-431

[18] Gaeta A L, Griffith A G, Cardenas J, et al. 2017 Low-loss silicon platform for broadband mid-infrared photonics *Optica* **4** 707-712

[19] Talebifard S, Schmidt S. Wei S, et al. 2017 Optimized sensitivity of Silicon-on-Insulator (SOI) strip waveguide resonator sensor Biomedical Optics Express, 2017, 8(2): 500-511.